\documentclass[12pt]{article}
\usepackage{latexsym}
\usepackage{amsfonts}
\usepackage{amssymb}
\usepackage[cp1250]{inputenc}

\title{Elementary proof of the bound on the speed of quantum evolution}
\author{Piotr Kosi\'nski\thanks{supported by the {\L}\'od\'z University Grants $N^0$\ 690 and 795 } ,  
 Magdalena Zych$^*$\\  
Department of Theoretical Physics II,\\Institute of Physics,
University of {\L}\'od\'z, \\
ul. Pomorska 149/153, 90 - 236 {\L}\'od\'z, Poland.}
\date{}

\begin{document}
\maketitle
\begin{abstract}
An elementary proof is given of the bound on "orthogonalization time": \\ 
$t_0\geq \frac{\pi  \hbar}{2\Delta E}$.
\end{abstract}
\newpage
In many problems of quantum theory (like, for example, quantum computing \cite{b1}$\div $\ \cite{b4} or 
fidelity between two quantum states \cite{b5}, \cite{b6}) it appears important to estimate speed of quantum evolution.

 An interesting measure of evolution speed is provided by the minimum 
time $t_0$\ required for the state to be transformed into an orthogonal (i.e. distinguishable) state. 
The basic estimate concerning $t_0$\ is given by the inequality
\begin{eqnarray}
t_0\geq \frac{\pi \hbar}{2\Delta E} \label{w1}
\end{eqnarray}
which has been derived and studied by many authors \cite{b7} $\div $\ \cite{b13}. This bound, in terms of energy dispersion 
$\Delta E$\ of initial state, is very simple and natural (in particular, $\Delta E=0$\ implies $t_0=\infty $\ as it should be 
 since the initial state 
is then an energy eigenstate). It has been generalized in various directions \cite{b14}, \cite{b15}, \cite{b5}; also, a 
beautiful geometric interpretation in terms of Fubini - Study metric was given \cite{b16} (see also \cite{b17}) and 
the intelligent states saturating (\ref{w1}) were found \cite{b18}. \\

Quite unexpectedly, few years ago Margolus and Levitin \cite{b1} derived a new bound of the form 
\begin{eqnarray}
t_0\geq \frac{\pi \hbar}{2(E-E_0)}    \label{w2}
\end{eqnarray}
 valid for Hamiltonians bounded from below; here $E_0$\ is the lowest energy while $E$\ is the expectation value of the 
Hamiltonian. They were able to show that, for a large class of states, eq. (\ref{w2}) provides a more optimal bound than 
eq. (\ref{w1}) (on the other hand, for energy eigenstates, except the lowest one, (\ref{w2}) is useless). The intelligent 
states for the 
inequality (\ref{w2}) were found in Refs. \cite{b19}, \cite{b20}.\\

While the standard proof of the bound (\ref{w1}) is based on Heisenberg equations of motion and uncertainty principle
 ( see, however, \cite{b12}), Margolus - 
Levitin derivation of the new bound (\ref{w2}) is surprisingly elementary; moreover, the corresponding intelligent states 
can be easily found \cite{b20}. \\

The question arises whether the bound (\ref{w1}) can be derived along the same lines. The aim of the present note is to 
provide the positive answer to this question. We shall show that (\ref{w1}) holds provided the Hamiltonian $H$\ is selfadjoint
and the initial state belongs to its domain. No further restrictions on the properties of $H$\ are necessary;
in particular, the spectrum may include both discrete and continuous parts and may extend to infinity in both directions.

Let us first sketch a generalization of the elegant approach of Ref.\cite{b1}. We assume for simplicity that the spectrum of $H$\
is purely discrete; the general case is briefly discussed in the final part of the paper.

 Let $\{\mid n\rangle \}$\ be the basis consisting of eigenstates of the Hamiltonian $H$,
\begin{eqnarray}
H\mid n\rangle=E_n\mid n\rangle  \label{w3}
\end{eqnarray}
and let
\begin{eqnarray}
\mid \Psi (0)\rangle=\sum\limits_nc_n\mid n\rangle  \label{w4}
\end{eqnarray}
be some initial state. Then
\begin{eqnarray}
\left\langle\Psi (0)\mid \Psi (t)\right\rangle=\sum\limits_n\mid c_n\mid ^2e^{\frac{-iE_n}{\hbar}t}=
\left\langle cos(\frac{Ht}{\hbar})
\right\rangle_
0-i\left\langle sin(\frac{Ht}{\hbar})\right\rangle_0  \label{w5}
\end{eqnarray}
here $\langle f(H)\rangle_0\equiv \sum\limits_nf(E_n)\mid c_n\mid ^2$\ denotes the average with respect to 
the initial state. \\
Now, due to $\langle \Psi (0)\mid \Psi (t_0)\rangle=0$\ one obtains
\begin{eqnarray}
\left\langle cos (\frac{Ht_0}{\hbar})\right\rangle_0=0,\;\;\; 
\left\langle sin (\frac{Ht_0}{\hbar})\right\rangle_0=0  \label{w6}
\end{eqnarray}
or
\begin{eqnarray}
\left\langle A cos (\frac{Ht_0}{\hbar}+\alpha ) \right\rangle_0=0  \label{w7}
\end{eqnarray}
for arbitrary constants $A,\alpha $.\\

Consider now an inequality of the form
\begin{eqnarray}
f(x)\geq A cos (x+\alpha )  \label{w8}
\end{eqnarray}
which is assumed to hold for $- \infty <x<\infty$\ or $0\leq x\leq \infty$\ if the spectrum of $H$\ extends in both 
directions or is nonnegative, respectively. Then
\begin{eqnarray}
\left\langle f(\frac{Ht}{\hbar})\right\rangle_0\geq \left\langle A cos (\frac{Ht}{\hbar}+\alpha ) \right\rangle_0  \label{w9}
\end{eqnarray}
provided the left-hand side is well defined ( i.e. average exists). Now, due to eq. (\ref{w7}), 
\begin{eqnarray}
\left\langle f(\frac{Ht_0}{\hbar})\right\rangle_0\geq 0  \label{w10}
\end{eqnarray}
The above inequality imposes certain restrictions on $t_0$. By a judicious choice of $f(x)$\ one can learn something 
interesting about $t_0$. For example, the bound (\ref{w2}) is obtained taking the optimal inequality (\ref{w8}) in the class 
of linear functions $f(x)$\ (in this case we have to restrict the range of $x$\ to positive semiaxis). \\
Let us now consider (\ref{w8}) in the class of quadratic functions $f(x)$\ and $-\infty<x<\infty$. It is an elementary task to check that 
the optimal inequality reads now
\begin{eqnarray}
(x+\alpha )^2-\frac{\pi }{4}\geq -\pi  cos (x+\alpha )  \label{w11}
\end{eqnarray}
By assumption, $\mid \Psi (0)\rangle$\ belongs to the domain of $H$\ and both $\langle H \rangle_0$\ and $\langle H^2 \rangle_0$\
are well defined \cite{b21}.
Eq. (\ref{w10}) takes now the form
\begin{eqnarray}
\frac{\langle H^2\rangle_0}{\hbar^2}t_0^2+\frac{2\alpha \langle H\rangle_0}{\hbar}t_0+(\alpha ^2-\frac{\pi ^2}{4})\geq 0  
\label{w12}
\end{eqnarray}
which implies that $t_0$\ lies outside the open interval
\begin{eqnarray}
 \Delta _{\alpha }  
\equiv\left(\frac{-2\alpha \langle H \rangle_0-\sqrt{\pi ^2\langle H^2 \rangle_0-4\alpha ^2\Delta E^2_0}}
{\frac{2\langle  
H^2\rangle_0}{\hbar}}\;,\;\frac{-2\alpha \langle H\rangle_0+\sqrt{\pi ^2\langle H^2\rangle_0-4\alpha ^2\Delta E^2_0}}{\frac
 {2\langle H^2\rangle_0}{\hbar}}\right) \label{w13}
\end{eqnarray}
where $\Delta E^2_0\equiv \langle H^2\rangle_0 - \langle H\rangle^2_0$.
It follows from eq.(\ref{w13}) that $\Delta _\alpha $\ is nonempty provided $\alpha $\ belongs to the open interval
\begin{eqnarray}
\Omega  \equiv \left ( \frac{-\pi \sqrt{\langle H^2\rangle_0}}{2\Delta E_0}\;, \; \frac{\pi \sqrt{\langle H^2\rangle}}                                                                                                                             
{2\Delta E_0} \right )  \label{w14}
\end{eqnarray}
So, finally, we obtain
\begin{eqnarray}
t_0 \notin \bigcup _{\alpha \in \Omega } \Delta _\alpha =\left(\frac{-\pi \hbar}{2\Delta E_0}\;, \; \frac{\pi \hbar}{2\Delta E_0}\right) 
\label{w15}
\end{eqnarray}
which implies (\ref{w1}).\\

In order to find intelligent states for the bound (\ref{w1}) we define
\begin{eqnarray}
\gamma  _\alpha (x)\equiv (x+\alpha )^2-\frac{\pi ^2}{4}+\pi cos (x+\alpha )  \label{w16}
\end{eqnarray}
Then
\begin{eqnarray}
\gamma  _\alpha (x)\geq 0  \label{w17}
\end{eqnarray}
and $\gamma  _\alpha (x)=0$\ if and only if $x=-\alpha \pm \frac{\pi }{2}$. \\
Assuming $t_0=\frac{\pi \hbar}{2\Delta E_0}$\ we find from (\ref{w12}) and (\ref{w16})
\begin{eqnarray}
\left\langle \gamma  _\alpha (\frac{Ht_0}{\hbar})\right\rangle_0=0 \;\;\; for \;\;\; \alpha =\frac{-\pi \langle H \rangle_0}
{2\Delta E_0} 
\label{w18}
\end{eqnarray}
Now, due to (\ref{w17}), eq.(\ref{w18}) implies $c_n\neq 0$\ only if $\frac{E_nt_0}{\hbar}=\frac{\pi \langle H \rangle_0}{2
\Delta E_0}\pm \frac{\pi }{2}$. Therefore, $c_n\neq 0$\ for at most two levels and $E_{n_1}=\langle H \rangle_0+\Delta E_0, \;\;
E_{n_2}=\langle H \rangle_0-\Delta E_0$\ which holds provided $\mid c_{n_1}\mid ^2=\mid c_{n_2}\mid ^2=\frac{1}{2}$. Therefore, the 
intelligent states are of the form \cite{b18}
\begin{eqnarray}
\mid \chi \rangle =c_1\mid n_1\rangle +c_2\mid n_2\rangle, \;\;\; \mid c_1\mid ^2=\mid c_2\mid ^2=\frac{1}{2}   \label{w19}
\end{eqnarray}

Finally, let us briefly discussed the general case when no assumption concerning the spectrum of $H$\ is made. Spectral theorem
\cite{b21} allows us to write
\begin{eqnarray}
\langle \Psi (0)\mid \Psi (t) \rangle = \langle \Psi (0)\mid e^{\frac{-it}{\hbar}H}\mid \Psi (0) \rangle =
\int e^{\frac{-iEt}{\hbar}} d\langle \Psi (0)\mid P_E\mid \Psi (0) \rangle  \label{w20}
\end{eqnarray}
where $P_E$\ is a spectral measure for energy. By assumption $\mid \Psi (0) \rangle$\ belongs to the domain of $H$\ 
which implies  \cite{b21} 
\begin{eqnarray}
\int E^2 d\langle \Psi (0)\mid P_E\mid \Psi (0) \rangle < \infty  \label{w21}
\end{eqnarray}
Therefore, $\gamma _\alpha (\frac{Et}{\hbar})$\ is integrable and
\begin{eqnarray}
\int \gamma _\alpha (\frac{Et}{\hbar}) d\langle \Psi (0)\mid P_E\mid \Psi (0) \rangle \geq  0 \label{w22}
\end{eqnarray}
which again leads to the estimate (\ref{w1}).

\end{document}